\documentstyle[preprint,aps,eqsecnum]{revtex}
\begin{document}
\draft
\title{Low-temperature dynamical simulation of spin-boson systems}
\author{Reinhold Egger\thanks{Present address:
Fakult\"at f\"ur Physik, Universit\"at Freiburg,
Hermann--Herder--Str.~3, D-79104 Freiburg, Germany}
  and C.H. Mak}
\address{Department of Chemistry, University of Southern California\\
Los Angeles, California 90089-0482, USA}
\maketitle
\begin{abstract}
The dynamics of spin-boson systems at very low
temperatures has been studied
using a real-time path-integral simulation technique
which combines a stochastic Monte Carlo sampling
over the quantum fluctuations with an exact treatment of the quasiclassical
degrees of freedoms. To a large degree, this special technique circumvents the
dynamical sign problem and allows
the dynamics to be studied directly up to long
real times in a numerically exact manner.  This method has been applied
to two important problems: (1)
crossover from nonadiabatic to adiabatic behavior in
electron transfer reactions,
(2) the zero-temperature dynamics in the antiferromagnetic Kondo region
 $1/2<K<1$ where $K$ is Kondo's parameter.
 \end{abstract}

\narrowtext
\section{INTRODUCTION}

Spin-boson systems provide archetypical models for
many low-temperature
dissipative quantum tunneling systems
 \cite{weiss93,caldeira,leggett87,weiss87}. Applications
include diverse problems such as
the observability of macroscopic quantum
coherence in a superconducting quantum interference device
(SQUID) \cite{weiss87}, interstitial tunneling of light
particles such as hydrogen in metals \cite{wipf},
and many others discussed in Ref.\cite{weiss93}.
Two other examples are of special importance to this work.
First, the so-called Kondo problem concerns
localized spin impurities in nonmagnetic metals \cite{tsvelick}.
The second comes from the realm of chemical physics ---
in certain parameter regions,
the spin-boson model provides a generalization of the Marcus
model for electron transfer reactions
\cite{marcus85,ulstrup,chandler91}. The
wide range of applicability of the spin-boson model
stems from its apparent
generality -- by coupling two exactly solvable
models bilinearly, namely a two-state system (spin)
and an infinite-dimensional harmonic oscillator (boson) bath,
one obtains a nontrivial description for many dissipative systems\cite{foot0}.

The dynamics of the spin-boson system has been widely studied
using the Feynman-Vernon influence functional
 method\cite{feyn63,chandler81}, mostly in conjunction
with instanton techniques \cite{leggett87}. Much of the physics
of this model has been unraveled by analytical methods,
though an exact solution is not possible (except for some special
parameter values).
 One particularly useful analytical approximation is the noninteracting
blip approximation (NIBA) \cite{leggett87}. Although the NIBA has been
rather successful, there
are important regions in parameter space where this approximation
is expected to fail.

Very recently, numerical
techniques for computing the dynamics of dissipative two-state systems
have been developed\cite{mak91,mak92,egger92,mak93a}.
These real-time quantum Monte Carlo (QMC) simulations have confirmed the
NIBA predictions quantitatively in many cases,
 and in addition have provided information
in regions where the NIBA breaks down.
However, due to the fundamental
{\em dynamical sign problem} inherent in these
 calculations \cite{filinov,doll,makri,gubernatis,cline,mason},
 the numerical computations
are restricted in their accessible range of real times.
The dynamical sign problem arises because
 at long times a large number of interfering paths contribute,
leading to a very small signal-to-noise ratio.
 In effect, the simulation becomes
unstable. Consequently, even numerically exact QMC methods have not
been able to resolve many important questions concerning the behavior
of the spin-boson system, especially at low temperatures and long times.

It is important to point out that many other approaches to dealing
with the dynamical sign problem have been developed
 during the last decade.  For example, methods
based on related filtering techniques
\cite{filinov,doll}, optimized reference systems \cite{makri}, or
analytic continuation procedures following a
conventional imaginary-time QMC simulation
\cite{gubernatis} have been proposed.
Similar to the one described in this work, the first two methods
attempt a direct simulation in real time.
While we restrict our attention to {\em discrete} (tight-binding)
systems which is adequate for many problems in low-temperature physics,
these previous methods are designed for extended systems.
On the other hand,
the third method makes use of an imaginary-time simulation that
has no dynamical sign problem to obtain numerical data which are then
analytically continued to real times.
To date this procedure has only be used to
find (e.g., electronic) spectral densities.  While real-time
information follows from the knowledge of these densities, such
calculations have yet to be performed and hence the effectiveness of
this approach to real-time dynamics is untested.
We also point out that a similar numerical problem
exists for QMC simulations of fermionic
many-body systems\cite{loh}. This {\em fermion
sign problem} has a different origin from the one
dealt with here. It
is due to the antisymmetry of fermionic wavefunctions,
and makes even imaginary-time simulations problematic.
If simulated in imaginary time, however, the spin-boson
model does not pose any sign problem.

In this paper we propose a new simulation
method to study the spin-boson dynamics
numerically. Like our earlier approaches,
this technique is based on a discretized
path integral representation of the dynamical
quantities of interest, but differs from them
in that it exploits the symmetry properties of the influence functional.
This allows for an exact treatment of the numerically problematic
quasiclassical paths, and one is left with a stochastic Monte Carlo sampling
of the quantum fluctuations alone.
Related methods have been used previously to study quantum Brownian
motion \cite{cline,mason},
and we have used a similar algorithm
to study the primary electron transfer steps in
the bacterial photosynthetic reaction center\cite{egger93}. Furthermore,
this idea of exploiting the symmetries of the influence
functional has also led to an efficient simulation method
for computing the mobility and diffusion coefficient of
a dissipative particle in an infinite tight-binding lattice \cite{mak93b}.
The algorithm for this diffusion problem,
 however, is rather different from the one employed in
the case of a system with a small number of tight-binding
states. In this article, we provide
a detailed description of our simulation method for the dissipative
two-state system; the necessary generalizations for the
case of more than two states should be obvious.

We apply this special method to the spin-boson model in the low-temperature
and the small bath cutoff regions.
The accuracy of the NIBA in some other parameter regions
 has been confirmed by our earlier simulations already, and the numerical
results presented in this work focus on these
previously inaccessible regions.
Since the present algorithm is more powerful than previous methods,
we are able to study longer times and much lower temperatures.

Data are presented for two practical problems of current interest:
(1) Electron transfer in a condensed phase
environment, where we have examined the transition from
nonadiabatic to adiabatic behavior for both the high-temperature and
the low-temperature case. The high-temperature results should
coincide with classical Marcus theory, whereas for low temperatures
we would expect quantum effects commonly
attributed to nuclear tunneling \cite{marcus85}.
 (2) The dissipative two-state system is also significant for
 the antiferromagnetic Kondo problem of
a localized impurity embedded  in a nonmagnetic metal \cite{tsvelick}.
The anisotropic Kondo Hamiltonian is related to the spin-boson
model in a certain parameter region, and we have studied the most interesting
antiferromagnetic case $1/2<K<1$, where $K$ is Kondo's parameter,
at zero temperature.  An important question to be
addressed in this parameter region relates to the destruction
of quantum coherence. For $K<1/2$, one can observe
quantum coherence (oscillatory behavior) in the
zero-temperature dynamics of the dissipative system.
The NIBA predicts  such coherent behavior to be completely destroyed
 for $K>1/2$. Unfortunately, the justification for NIBA is suspect
in this special parameter region, and our method offers an
unique way of studying this region.

In Section II we
present our simulation method for the dissipative two-state system.
The results for
the crossover between nonadiabatic and adiabatic electron transfer
are discussed in Section III, followed by a discussion of
the dynamics in the  antiferromagnetic Kondo region at very low
temperatures. Some final remarks and conclusions
are given in Section IV.

\section{SIMULATION METHOD}

In the following we present a dynamical simulation technique
for dissipative tight-binding models with a finite number of states.
For simplicity, the discussion is restricted
to the case of two states,
which leads to the often-studied spin-boson Hamiltonian \cite{leggett87}
\begin{eqnarray} \label{spbos}
H &=& H_0+H_B + H_I \\
&=& - (\hbar\Delta/2)\, \sigma_x + (\hbar\epsilon/2)\, \sigma_z + \sum_\alpha
\left[ \frac{p_\alpha^2}{2m_\alpha} +
{\textstyle \frac{1}{2}}  m_\alpha \omega_\alpha^2
\left(x_\alpha - \frac{C_\alpha}{m_\alpha \omega_\alpha^2} \sigma_z\right)^2
\right] \;.
\nonumber
\end{eqnarray}
The parameters in the free Hamiltonian $H_0$ describing the isolated
two-state system are the tunnel matrix element $\Delta$ --- in the parlance
of electron transfer theory, $\Delta/2$ is the electronic coupling
 between the two different redox sites,
 the external bias  $\epsilon$ corresponds to an asymmetry
between the two localized energy levels, and $\sigma_x$ and $\sigma_z$ are
the usual Pauli matrices. The bath is described by
harmonic oscillators $\{x_\alpha\}$
which are bilinearly coupled to the spin operator $\sigma_z$. This
type of coupling is reasonable for the problems considered
in this work (and for many others); for a justification,
see, e.g., Refs.\cite{weiss93,caldeira,leggett87}.

Within this model, the bath parameters enter only via
a single function called the spectral density
\begin{equation}
J(\omega)=\frac{\pi}{2}
\sum_\alpha \frac{C^2_\alpha}{m_\alpha\omega_\alpha}\,
\delta (\omega-\omega_\alpha) \;,
\end{equation}
which should have a continuous form in the limit of infinitely many bath
oscillators. The spectral density then determines the
 bath correlation function \cite{chandler91}
\begin{equation}\label{bathcorr}
L(z) = \frac{a^2}{\pi \hbar} \int_0^\infty d\omega \, J(\omega)
\, \frac{\cosh[\omega(\hbar\beta/2-iz)]}{\sinh[\omega\hbar\beta/2]}\;,
\end{equation}
which is defined for complex-valued times
$z=t-i\tau$.
The distance between the localized states is given by a lengthscale $a$,
 and $\beta = 1/k_{\rm B} T$.
 Here, we limit our attention to the
case of an Ohmic spectral density which has the form
\begin{equation}\label{ohmic}
J(\omega) = (2\pi\hbar K/a^2)\, \omega \, e^{-\omega/\omega_{\rm c}}\;.
\end{equation}
This spectral density has a characteristic low-frequency behavior
$J(\omega) \sim \eta \omega$, where $\eta$ is the usual Ohmic viscosity.
 The system-bath coupling strength  is
measured in terms of the dimensionless Kondo parameter $K$.  The
timescale distribution of bath motions is described by
 a cutoff frequency $\omega_{\rm c}$. For many
problems in low-temperature physics, this cutoff
frequency is taken to be the largest frequency scale in the problem.
In the case of electron transfer, the same spectral density with
some intermediate value for $\omega_{\rm c}$ is most appropriate for a
realistic description of many polar solvents\cite{egger93b}.

Two dynamical quantities of interest for this model are the
symmetrized time correlation function
\begin{eqnarray}
C(t) &=& \mbox{Re} \, \langle \sigma_z(0) \sigma_z (t)\rangle \nonumber\\
&=& Z^{-1} \mbox{ReTr} \left( e^{-\beta H} \sigma_z e^{iHt/\hbar}
 \sigma_z e^{-iHt/\hbar} \right)
\label{ctt}
\end{eqnarray}
with $Z=\mbox{Tr}\, e^{-\beta H}$, and the
time-dependent occupation probabilities for the two sites $P_+(t)$ and
$P_-(t)$ which can be expressed in terms of a single function
\begin{equation}\label{ptt}
P(t) = P_+(t) - P_-(t) = \langle e^{iHt/\hbar}\sigma_z e^{-iHt/\hbar}\rangle
\end{equation}
with the initial condition $P(0)=1$.
{}From a comparison of Eqs.(\ref{ctt}) and (\ref{ptt}) we observe that
the two quantities differ only by the
way the system is prepared initially. For $P(t)$, the
system is held fixed in the $+$ state until $t=0$
 with the bath being
unobserved (factorized initial condition).
 On the other hand, the more realistic situation for many
experiments is represented by equilibrium states of the total system
at $t=0$ as described by $C(t)$.

To numerically compute the two-state dynamics, we employ
a discretized path-integral representation of the dynamical quantities
\cite{chandler81,mak91,mak92,egger92}. The correlation function
$\langle \sigma_z(0) \sigma_z(t) \rangle$ can be regarded
as the probability amplitude for a sequence of steps in the complex-time
plane. In particular, one propagates along the Kadanoff-Baym contour
$\gamma$: $z=0\to t \to 0 \to -i\hbar\beta$, and measures
$\sigma_z$ at $z=0$ and $z=t$. Of course, there are several other
possible choices of this contour  due to the cyclic structure of
the trace.

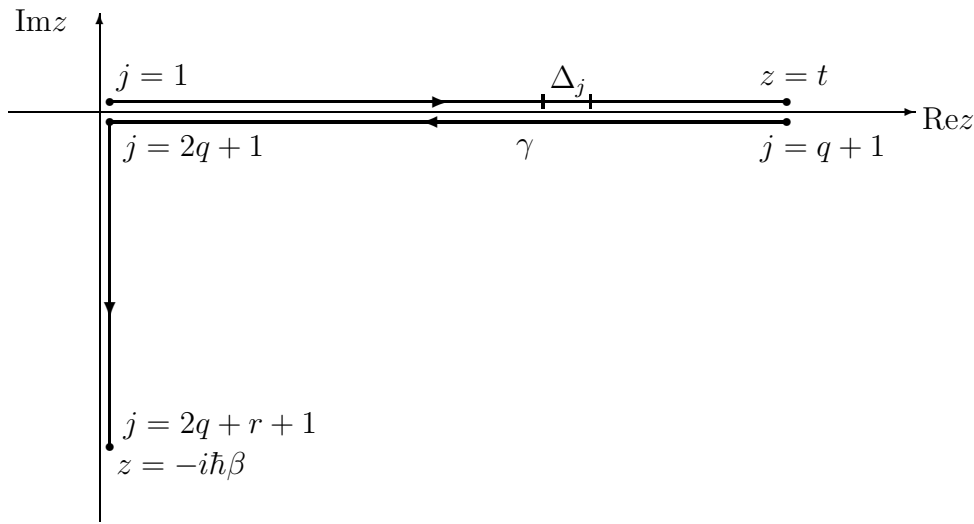
\begin{figure}[t]
\unitlength0.9cm
\begin{picture}(18,8.5)
\thinlines
\put(2.85,1.1){\line(0,1){7.1}}
\put(1.5,7.15){\line(1,0){13}}
\put(2.85,8.2){\vector(0,1){.4}}
\put(14.5,7.15){\vector(1,0){.4}}
\put(15,6.9){Re$z$}
\put(1.6,8.35){Im$z$}
\thicklines
\put(3,7.3){\line(1,0){10}}
\put(3,7){\line(1,0){10}}
\put(13,7){\circle*{0.12}}
\put(13,7.3){\circle*{0.12}}
\put(12.6,7.55){$z=t$}
\put(12.61,6.5){$j=q+1$}
\put(3,7){\circle*{0.12}}
\put(3,7.3){\circle*{0.12}}
\put(3.1,7.55){$j=1$}
\put(3.2,6.5){$j=2q+1$}
\put(8,7){\vector(-1,0){0.4}}
\put(7.6,7.3){\vector(1,0){0.4}}
\put(9,6.55){$\gamma$}
\put(9.4,7.2){\line(0,1){0.2}}
\put(10.1,7.2){\line(0,1){0.2}}
\put(9.5,7.5){$\Delta_j$}
\put(3,2.2){\line(0,1){4.8}}
\put(3,2.2){\circle*{0.12}}
\put(3,4.5){\vector(0,-1){0.4}}
\put(3.2,2.4){$j=2q+r+1$}
\put(3.1,1.8){$z=-i\hbar\beta$}
\end{picture}
\caption[]{\label{fig0} Discretization of the Kadanoff-Baym contour
$\gamma$ in the complex-time plane. }
\end{figure}

To parametrize the paths, we
use $q$ uniformly spaced
discrete points for each of the two real-time paths and
$r$ points for the imaginary-time path (see Fig.\ref{fig0}).
Hence, there are $N=2q+r$ points in total.
 The time discretizations are
\begin{equation}\label{timdis}
\Delta_j = \left\{ \begin{array}{c@{\quad,\quad}l}
t/q & 1\leq j \leq q \\
-t/q & q+1 \leq j\leq 2q \\
-i\hbar\beta/r & 2q+1 \leq j \leq 2q+r \equiv N \;,
\end{array} \right.
\end{equation}
and the complex time after $i$  steps is $z_i=\sum_{j=1}^{i-1} \Delta_j$.
The construction of the
discretized path integral then proceeds  by inserting complete sets
\begin{eqnarray} \nonumber
1_i &=& \sum_{\sigma_i=\pm 1} \int \prod_\alpha dx_{\alpha,i}
\,|\sigma_i,\{x_{\alpha,i}\}\rangle\langle \sigma_i,\{x_{\alpha,i}\}|
\\ &=& \int d{\bf r}_i\; |{\bf r}_i \rangle \langle {\bf r}_i |
\label{one-repr}
\end{eqnarray}
at each discretization point $z_i$ ($i=2,\ldots,N$).
The vector ${\bf r}_i$ represents the state $\sigma_i=\pm 1$ of the
two-level system as well as
the environmental degrees of freedom $\{x_{\alpha,i}\}$.

To disentangle the short-time propagator, we
use a (symmetrized) Trotter formula,
\begin{equation}
\exp(-iH\Delta_j/\hbar) = \exp(-iH' \Delta_j/2\hbar)
\exp(-iH_0 \Delta_j/\hbar) \exp(-iH'\Delta_j/2\hbar)
+ {\cal O}(\Delta_j^3 [H_0,H^\prime])\;,
\end{equation}
where $H'=H-H_0$ is diagonal
in the representation (\ref{one-repr}). The free part $H_0$ of
the Hamiltonian (\ref{spbos}) leads to  the short-time propagator
\begin{equation}\label{freematel}
 K(\sigma_j,\sigma_{j+1}) =
\langle \sigma_{j+1}| \exp(-i\Delta_j H_0/\hbar) | \sigma_j\rangle\;
\end{equation}
 which can be evaluated exactly. Thereby we arrive at
\begin{equation} \label{disc}
\langle \sigma_z(0) \sigma_z(t') \rangle = \frac{
\int d^N {\bf r}\; e^{S[{\bf r}_1,\ldots,{\bf r}_N]}
 \sigma'  \sigma^{\prime\prime}}{
\int d^N {\bf r}\; e^{S[{\bf r}_1,\ldots,{\bf r}_N]} }\;,
\end{equation}
which converges to the true path integral as
the number of discretizations $N\to \infty$. The
discretized action $S[{\bf r}_1,\ldots,{\bf r}_N]$ is a complex-valued
sum of the actions picked up in separate parts of the contour $\gamma$.
To compute the correlation function in Eq.(\ref{disc}), one
can average over all pairs of spins $\{\sigma',\sigma^{\prime\prime}\}$
separated by a time $t'$ along the contour $\gamma$. This allows us
to compute the dynamical quantities for all times $t_k= kt/q$
(where $k=0,\ldots, q$)
 from one single Metropolis trajectory. Because
of the cyclic structure of the trace in Eq.(\ref{ctt}), ${\bf r}_{N+1}
\equiv {\bf r}_1$.

Since the bath is made up of harmonic oscillators,
one can integrate out
the environmental degrees of freedom analytically.
After performing this integration,
the bath-plus-coupling part $H'$ of the
Hamiltonian leads to an {\em influence functional} $\Phi[\sigma]$
in terms of the spins
$\{ \sigma_i\}$ alone\cite{feyn63,chandler81}. As a result, the
correlation function takes the form
\begin{equation}\label{co}
\langle \sigma_z(0) \sigma_z(t') \rangle = \frac{1}{Z}
\sum_{\{\sigma\}} \exp\left( -\Phi[\sigma] + \sum_i \ln K(\sigma_i,
\sigma_{i+1}) \right) \sigma' \sigma^{\prime\prime}\;,
\end{equation}
where $Z=\sum_{\{\sigma\}} \exp(\ldots)$ (the exponent will
be referred to as ``the action'' henceforth).
In the continuum limit $N\to \infty$,
 the nonlocal influence functional is given
in terms of the bath correlation function $L(z)$ introduced in
Eq.(\ref{bathcorr}), and one finds with $\sigma_i \to
\sigma(z)$  \cite{grabert88}
\begin{equation}
\Phi[\sigma] = \int_\gamma dz \int_{z'<z} dz'  \; \sigma(z)
L(z-z') \sigma(z')/4 \;.
\end{equation}
The integrations in the complex-time plane are ordered
along the Kadanoff-Baym contour $\gamma$.
Note that $L(z)$ fulfills the important symmetry relation
\begin{equation}\label{symm}
L(z-i\hbar\beta) = L(-z)\;,
\end{equation}
which implies certain symmetry properties of the influence functional.

In discretized form, the influence functional is given by
\cite{chandler81}
\begin{equation}\label{infl}
\Phi[\sigma]  = {\textstyle \frac{1}{2}}
\sum_{j,k=1}^N \sigma_j L_{jk} \sigma_k/4 \;,
\end{equation}
with the complex-valued influence matrix $L_{jk}$.
We observe from
  Eqs.(\ref{co}) and (\ref{infl}) that the
(discretized) spin-boson problem is
isomorphic to a classical one-dimensional
Ising model with long-range complex-valued interactions
\cite{chandler81,mak91}.  The
influence matrix is the average value of the influence functional interaction
between two points $z_j$ and $z_k$
\begin{equation}
\label{defll}
L_{jk} = L_{kj} = \int_{C_j} dz'_j \int_{C_k} dz'_k\; L(z'_j - z'_k)
\qquad (\mbox{for} \; j>k) \;,
\end{equation}
where $C_i$ is one discretization on the contour centered at the point
$z_i$, i.e.,
\[
\{ z\in C_i | z_i - \Delta_{i-1}/2 < z < z_i + \Delta_i/2 \}\;.
\]
The remaining time integrations in Eq.(\ref{defll}) can be carried out easily,
and one finds with $\Delta_{jk} = z_j - z_k$
\begin{eqnarray}  \nonumber
L_{jk} & = &Q(\Delta_{jk} + (\Delta_j+\Delta_{k-1})/2)
          + Q(\Delta_{jk} + (-\Delta_{j-1} - \Delta_k)/2) \\
   &-& Q(\Delta_{jk} + (-\Delta_{j-1} + \Delta_{k-1})/2)
     - Q(\Delta_{jk} + (\Delta_j - \Delta_k)/2) \;, \label{lmatr}
\end{eqnarray}
where $Q(z)$ is the twice-integrated bath correlation function with
$Q(0)=0$, i.e., $d^2 Q(z)/dz^2 = L(z)$. Of course, this function exhibits
the same symmetry property (\ref{symm}).
Finally, the diagonal elements
are given by
\[
L_{jj} = 2Q((\Delta_{j-1}+\Delta_j)/2) \;.
\]
A detailed derivation of the influence matrix can be found in
Refs.\cite{egger92,egger93}.

Since the action for each spin configuration (``a path'') is complex-valued,
a stochastic Monte Carlo evaluation of the resulting isomorphic Ising chain
is faced with the dynamical
sign problem. In the past, we have partially
circumvented this problem either  by transforming to a continuous
spin representation and applying stationary-phase Monte Carlo (SPMC)
methods \cite{mak91,egger92},
or  by introducing a local filtering function
in discrete state space\cite{mak92,mak93a}. The latter
 approach is related to ideas like the stationary-phase approximation and
allowed for a study of many phenomena on an intermediate timescale.

Here, we observe that
a much more efficient method
can be constructed when one takes into account
the symmetry properties of the influence matrix due to
  Eq.(\ref{symm}).  These symmetry
relations can be expressed mathematically as a set of index relations
 [we use $i,j,k=1,\ldots,q+1$, and  $n,m=2,\ldots,r$]
\begin{eqnarray*}
 L_{2q+2-i,2q+2-j} &=&  L^*_{ij} \\
 L_{2q+2-i,j} &=& -L^*_{ij} \qquad \mbox{for} \; j>i\\
 L_{2q+2-i,j} &=& -L_{ij} \qquad \mbox{for} \; i>j\\
L_{2q+2-i,i} &=& -\mbox{Re}\, L_{ii} \\
L_{2q+m,2q+2-j} &=& - L_{2q+m,j} \qquad \mbox{for}\; j>1\\
 L_{i,q+1} &=& L_{2q+2-i,q+1} =  L_{2q+m,q+1} = 0 \;,
\end{eqnarray*}
which can be proved easily from Eq.(\ref{lmatr}).

The benefit of exploiting these relations is realized upon
switching to a new spin representation.
To that purpose, we introduce the sum and difference coordinates of the
forward ($\sigma_j$) and backward  ($\sigma_j^\prime$)
real-time spin paths and rename the imaginary-time spins ($\bar{\sigma}_m$),
\begin{eqnarray}
\eta_j &=& (\sigma_j + \sigma'_j)/2  \nonumber\\
\xi_j &=& (\sigma_j - \sigma'_j)/2  \label{xidef}\\
\bar{\sigma}_m & = & \sigma_{2q+m}\;,\nonumber
\end{eqnarray}
where $\sigma'_j=\sigma_{2q+2-j}$ in the old notation ($j=1,\ldots,q+1)$.
 Thus we
first relabel the spins on the three pieces of the Kadanoff-Baym contour
and then form the said linear combinations.  A
physical meaning can be assigned to these new spins:
$\{\eta_j\}$ describe the propagation
along the diagonal of the reduced density matrix
and can thus be identified with
 {\em quasiclassical paths}, while
 $\{\xi_j\}$ describe the off-diagonaliticity of the reduced
density matrix and can be identified with {\em quantum fluctuations}
\cite{schmid82}. According to the definitions
(\ref{xidef}), the new spins can take on three
possible  values $\xi,\eta=-1,0,1$,
but they are not entirely independent because either
 $\xi_j$ or $\eta_j$ has to be 0 for the same $j$.

Written in terms of the new spins and exploiting the
index relations above, the influence functional takes the form
\begin{eqnarray}
\Phi[\xi,\eta,\bar{\sigma}] &=& {\textstyle \frac{1}{2}}
\sum_{m,n=2}^r \,\bar{\sigma}_m Y_{mn} \bar{\sigma}_n
+ {\textstyle \frac{1}{2}}
 \sum_{j,k=1}^q \, \xi_j \Lambda_{jk}\xi_k  \nonumber\\
&+& i\sum_{j>k=1}^q \, \xi_j   X_{jk} \eta_k +
\sum_{j=1}^q \sum_{m=2}^r \, \xi_j  Z_{jm} \bar{\sigma}_m
\label{inflma}
\\ &+& \eta_1 \left( \sum_{m=2}^r \bar{\sigma}_m \; [L_{2q+m,1}
+ L_{2q+m,2q+1}] \right)\;. \nonumber
\end{eqnarray}
 The elements of the matrices
appearing in Eq.(\ref{inflma}) are
\begin{eqnarray*}
Y_{mn} &=& L_{2q+m,2q+n}/4 \\
\Lambda_{jk} &=& \mbox{Re}\, L_{jk} \\
X_{jk} &=& \mbox{Im}\, L_{jk} \\
Z_{jm} &=& L_{j,2q+m}/2 \;.
\end{eqnarray*}
It is worth mentioning that
these matrices are {\em real-valued}
[with the exception of $Z_{jm}$]. The meaning of the five
terms  in Eq.(\ref{inflma}) is as follows. The first term describes
a self-interaction within the imaginary-time segment.
Due to the second term, the dissipative bath will
damp out the quantum fluctuations, and the system is
likely to be found in a diagonal state characterized by $\xi=0$.
 This point will later  be important
with regard to the choice of a suitable Monte Carlo weight.
The third term is a bilinear interaction between quasiclassical
paths and quantum fluctuations, and the fourth term
describes a similar interaction between the imaginary-time
spins and the quantum fluctuations. The last term is a preparation term,
coupling $\eta_1$ to the imaginary-time spins.
Remarkably, there is no self-interaction in the quasiclassical paths,
and they are coupled only linearly to other degrees of freedom.

This observation is crucial for our computational procedure, since
it allows for an exact treatment of the quasiclassical paths.
{\em For any given quantum fluctuation path,
  the path summation over all allowed quasiclassical paths
 can be carried out in an exact manner.}
 To elucidate this, we first examine the free
action due to $H_0$. The imaginary-time contribution can be put into
the matrix elements $Y_{mn}$ by simply adding $\frac12 \ln \tanh (\hbar
\beta\Delta/2r)$ to $Y_{m,m+1}$ and $Y_{m+1,m}$; in case an external
bias $\epsilon$ is present, the action has
to include an additional term
$(\hbar\beta\epsilon/2r) \sum_m \bar{\sigma}_m$ \cite{egger92}.
Regarding the real-time paths, we proceed in a different way.
 For the isolated two-state system, we have (in terms of the
original spins) a product of the form [cf.~Eq.(\ref{freematel})]
\begin{equation} \label{vgl}
\prod_{j=1}^{q} \, K(\sigma_j,\sigma_{j+1}) K^*(\sigma'_j,\sigma'_{j+1})\;.
\end{equation}
If we now switch to the $\{\xi,\eta\}$ representation and perform the
summation over all  $\eta$ spins (while keeping the $\xi$ configuration
frozen), we obtain a {\em matrix product} for Eq.(\ref{vgl}). Of course,
one has to account for the $\eta$-dependent terms in the influence
functional (\ref{inflma}) during this procedure. In the end, the
complex-valued contribution of all these terms for a given $\xi$
 configuration takes the form
\begin{equation} \label{factor}
{\cal J}[\xi] = \sum_{\eta_1=0,\pm 1} \sum_{\eta_{q+1}=0,\pm  1}
\langle \eta_1 | {\bf V}^{(1)} \ldots {\bf V}^{(q)} | \eta_{q+1} \rangle\;,
\end{equation}
where the $(3\times 3)$ matrices ${\bf V}^{(j)}[\xi]$
 are defined by\cite{foot1}
\begin{equation}
 \langle \eta_j | {\bf V}^{(j)} |\eta_{j+1} \rangle = [K\times K^*]
(\eta_j,\eta_{j+1},\xi_j,\xi_{j+1}) \exp \left[- i \eta_j \sum_{k>j}^q \,
X_{kj} \xi_k \right ]\;.
\end{equation}
Each of the matrices ${\bf V}^{(j)}$
depends on all $\xi_k$ spins with $k\geq j$; however, the ``free'' part
$K\times K^*$ is determined by $\xi_j$ and $\xi_{j+1}$ alone.
Clearly, ${\cal J}[\xi]$
 can be evaluated with a simple matrix multiplication routine,
leading to  a numerically exact and efficient treatment of the
quasiclassical paths. Note that the remaining part of the influence functional
 is real-valued
--- with the exception of the fourth term in Eq.(\ref{inflma}), which
is generally very small --- indicating that much of the dynamical
sign problem
has been relieved by treating the
numerically problematic quasiclassical
paths in an exact manner.

In effect, the factor ${\cal J}[\xi]$ contains all contributions from
$H_0$ and, in addition, the third and fifth term of the influence
functional (\ref{inflma}).
 The correlation function can thus be written as
\begin{eqnarray*}
\langle \sigma_z(0) \sigma_z(t') \rangle &=& \frac{1}{Z}
\sum_{\{\xi\}=-1,0,1} \sum_{\{\bar {\sigma}\} = \pm 1}
{\cal J}[\xi] \;
  \exp \Biggl(-{\textstyle \frac{1}{2}}
\sum_{m,n=2}^r \,\bar{\sigma}_m Y_{mn} \bar{\sigma}_n \\
&-& {\textstyle \frac{1}{2}}
 \sum_{j,k=1}^q \, \xi_j \Lambda_{jk}\xi_k
-\sum_{j=1}^q \sum_{m=2}^r \, \xi_j  Z_{jm} \bar{\sigma}_m\Biggr) \;
\sigma' \sigma^{\prime\prime}\;.
\end{eqnarray*}
Therefore we are left with the task of summing over the imaginary-time
spins $\{\bar{\sigma}_m\}$ and the quantum fluctuations $\{\xi_j\}$, which
is conveniently done via Monte Carlo (MC) sampling.
The  suitable MC weight for the imaginary-time spins is straightforward,
\[
{\cal P}_{\rm imag}[\bar{\sigma}] \sim \exp \left( -{\textstyle \frac{1}{2}}
\sum_{m,n} \bar{\sigma}_m Y_{mn} \bar{\sigma}_n
- \mbox{Re}\, \sum_{j,m} \xi_j Z_{jm} \bar{\sigma}_m \right)\;.
\]
Since the influence functional forces the quantum fluctuations to stay
near the diagonal of the density matrix, we first try to use the following
 MC weight for the quantum  fluctuations
 \[
{\cal P}_{\rm real}[\xi] \sim
\exp \left ( -
{\textstyle \frac{1}{2}} \sum_{j,k} \xi_j \Lambda_{jk}
\xi_k - \mbox{Re}\, \sum_{j,m} \xi_j Z_{jm} \bar{\sigma}_m \right) \;.
\]
The problem with this weight, however, arises for small system-bath
couplings where the damping of the quantum fluctuations becomes
weak. In this case, the importance sampling would become very
inefficient for small coupling $K$.
Hence, in a next step, we try the product
\[
   W[\xi] \sim {\cal  P}_{\rm real}[\xi] \times |{\cal J}[\xi]|\;,
\]
 where  ${\cal J}[\xi]$ has been defined in Eq.(\ref{factor}).
This weight function considers both the damping of the
quantum fluctuations due to the influence functional {\em and}\/
 the integrated-out quasiclassical paths.

Unfortunately, there is another problem with this weight, similar
to the case of the multistate algorithm \cite{mak93b,valleau}.
This problem arises since for correlation functions
like $\langle \sigma_z(0) \sigma_z(t) \rangle$
one has to compute the ratio of two quantities. It turns out that certain spin
configurations only contribute to the denominator but not to the
 numerator (and vice versa). Due to this exclusivity problem,
one will not be able to access all relevant spin configurations
$\{\xi\}$ contributing to the numerator when using $W[\xi]$
alone as the Monte Carlo weight. We can circumvent this problem
by observing that for the numerator, one has to compute
terms like [where $\alpha_1,\alpha_2 =\pm$]
\begin{eqnarray}
{\cal J}_{\alpha_1,\alpha_2}^{(k)} [\xi] &=&
\sum_{\eta_1=0,\pm 1} \sum_{\eta_{q+1}=0,\pm 1} \\ &&
\langle \eta_1 |H_{\alpha_1}(\xi_1) {\bf V}^{(1)}
 \ldots {\bf V}^{(k-1)} H_{\alpha_2}(\xi_k) {\bf V}^{(k)} \ldots
{\bf V}^{(q)} | \eta_{q+1} \rangle\;. \nonumber
\end{eqnarray}
The projection operators $H_+=(1+\sigma_z)/2$ and $H_-=(1-\sigma_z)/2$
onto the two spin values
 have the $\eta$-representation (for a given $\xi$)
\begin{eqnarray*}
H_+(\xi=0) = \left ( \begin{array}{ccc}
1 & 0 & 0 \\ 0& 0  & 0 \\ 0 & 0 & 0 \end{array}\right) &&\quad , \quad
H_-(\xi=0) = \left ( \begin{array}{ccc}
0 & 0 & 0 \\ 0& 0  & 0 \\ 0 & 0 & 1 \end{array}\right) \;,\\
H_\pm(\xi=\pm 1) &=& \left ( \begin{array}{ccc}
0& 0 & 0 \\ 0& 1/2  & 0 \\ 0 & 0 & 0 \end{array}\right) \;.
\end{eqnarray*}
Finally, an appropriate positive definite Monte Carlo weight function
can be constructed
\begin{equation}
\widetilde{W}[\xi] \sim {\cal P}_{\rm real}[\xi]
\left(\sum_{k=1}^q
\sum_{\alpha_1,\alpha_2=\pm } \left|{\cal J}_{\alpha_1,\alpha_2}^{(k)}
\right| \right)\;.
\end{equation}
Using $\widetilde{W}[\xi]$ as the weight allows us to carry out
an efficient Monte Carlo sampling of the $\xi$ spins.

Our QMC algorithm employs
single particle Metropolis moves as well as  moves  which allow kinks
to translate along the spin chain\cite{mak93a}. Single-particle moves
attempt to change one spin $\xi_k = -1,0,1$ to a new value, whereas
kink moves attempt to change two adjacent spins simultaneously.
The imaginary-time spins are sampled from
${\cal P}_{\rm imag}[\bar{\sigma}]$ using single particle moves, i.e.,
one tries to flip a single spin $\bar{\sigma}_m=\pm 1$.
During one MC pass, the single-particle moves are attempted
once for every spin, and the kink moves are attempted for
every pair of spins with $\xi_k \neq \xi_{k+1}$. Typical
acceptance ratios for these types of moves are $\approx 15\%$,
and we take samples separated by 5 MC passes. This ensures
that the MC samples are sufficiently uncorrelated, since
roughly half of the spins have been assigned new values
between two subsequent samples.
Numerical results are then obtained from several 10,000 samples,
with statistical errors always below $5\%$ for the data
reported here.

The calculations were carried out
on an IBM RISC 6000/580 workstation, at an average speed of
2 CPU hours per 10,000 samples
(for $q\approx 80, r=0$). As mentioned earlier,
the dynamical quantities of interest
can be sampled from a single Monte Carlo trajectory
for all times $t'\leq t$ since the remaining part of the Kadanoff-Baym
contour can be integrated out. Furthermore, the quantity
 $P(t)$ can be calculated using the same code
by simply removing all spins from the imaginary-time path,
i.e., by putting $r=0$. Finally, to ensure that the Trotter error is
sufficiently small, we have to keep the discretization numbers
$q$ and $r$ large enough. This is checked by systematically increasing
these numbers until convergence is reached. For all results presented
here, the Trotter error is negligible compared to the statistical errors
due to the
stochastic MC sampling (which are less than $5\%$).

We close this section with some remarks concerning the
relation between this method and our earlier techniques. The algorithm
presented here can be thought of as an optimized (but essentially
nonlocal) filtering method
for discrete-state systems as proposed by Mak \cite{mak92}. The
optimization is achieved by exploiting the underlying symmetries
of the influence functional.  This makes this method superior
in the sense that
we can study longer times with less computational effort. Since
the dynamical sign problem grows exponentially with increasing time, most of
the  results discussed in the next section
 cannot be obtained by any former technique.
An approximate measure for the performance of the different real-time
QMC algorithms for the spin-boson problem can be obtained from
the maximum real time $t_{\rm max}$ defined as the upper time limit
of the respective method. For times $t > t_{\rm max}$, the  large
statistical errors caused by the dynamical sign problem
(more than $\approx 20\%$) will render the simulation
results useless. This is quantified in
 Table I. The comparison with earlier methods demonstrates
 that the spin-boson dynamics can now be studied up to much longer
timescales despite the dynamical sign problem. The gain is most
 significant at very low temperatures, but is also
 important at higher temperatures.

\section{SIMULATION RESULTS}

In this section, we present dynamical simulation results
 for two different problems, namely
the crossover between nonadiabatic and adiabatic electron transfer
and the dynamics in the low-temperature antiferromagnetic
Kondo region. We will restrict our attention to the symmetric case
$\epsilon=0$ here, and we also consider only Ohmic spectral densities of
the form (\ref{ohmic}). For numerical results in other parameter
regions of the spin-boson  model, we refer to our previous  work
\cite{mak91,mak92,egger92,mak93a}.

\subsection{Crossover from nonadiabatic to adiabatic electron transfer}

The spin-boson system is an adequate model for many
electron transfer reactions in condensed phase systems \cite{chandler91}.
The electron transfer rate is in general determined not only by the
overlap of the electronic wavefunctions localized on the redox states
(which is proportional to the tunnel splitting $\Delta$), but also
by the properties of the solvent environment. For a charge transfer
to occur, a specific large-scale reorganization of the solvent is
required to achieve the resonance condition necessary for
electronic tunneling. Using linear response
theory for a description of the solvent modes
and a tight-binding model for the redox states, one arrives at
the spin-boson model. In our study, we have taken an Ohmic spectral
density (\ref{ohmic}) for the bath
with a finite cutoff frequency $\omega_{\rm c}$
which can be regarded as  a free parameter \cite{egger93b}.

The solvent is described by a continuous spectral density peaked around a
characteristic bath mode frequency $\omega_{\rm c}$, and
the classical reorganization energy corresponding to this bath is
$\hbar\lambda=2K\hbar\omega_{\rm c}$, where $K$ is Kondo's
parameter. For typical
electron transfer reactions, the reorganization energy fulfills the condition
$\lambda/\Delta\gg 1$. The nonadiabatic regime of electron transfer
is defined by small electronic couplings, $\Delta/\omega_{\rm c} \ll 1$,
whereas the adiabatic limit corresponds to electronic couplings of the order
of $\omega_{\rm c}$ (or larger). We note
that the nonadiabatic limit is realized in most biological and in
many chemical systems; nevertheless,  the adiabatic limit is also important
for a description of many chemical electron transfer  reactions.

The high-temperature limit of electron transfer
is well understood within the framework of classical Marcus theory
\cite{marcus85}. The rate can  be factored into an equilibrium Boltzmann
factor containing the activation free energy for the required global
bath fluctuation, and a Landau-Zener factor for the transition
 probability once this Landau-Zener region has been
reached\cite{ulstrup}.
 For symmetric electron transfer reactions, Marcus found
a bath activation energy $\hbar\lambda/4$, and combining this with
a conventional estimate for the Landau-Zener factor\cite{ulstrup}, one obtains
 a formula for the total (forward plus backward) rate \cite{garg}
\begin{equation}\label{clas}
\Gamma =
\sqrt{\frac{\pi \hbar\beta}{4 \lambda}}
\, \frac{\Delta^2}{1 + \Delta^2/\omega_{\rm b}\lambda}
\, \exp(-\beta \hbar\lambda/4)\;,
\end{equation}
which is valid for $\hbar\beta\omega_{\rm c}\ll 1$. The Landau-Zener
factor contains a  frequency scale $\omega_{\rm b} \approx \lambda$
reminiscent of transition state theory which is usually applied to
 the adiabatic limit\cite{chandler91}. Note that for
small electronic couplings, the rate $\sim \Delta^2$ which is just the
 golden rule behavior. In the adiabatic limit of large $\Delta$, however,
the rate becomes independent of $\Delta$, and the dynamics is totally
solvent-controlled.
We note that for the spectral density (\ref{ohmic}), one can obtain
analytical expressions in the nonadiabatic golden rule limit for both the
high- and low-temperature rate ($T'=\hbar\omega_{\rm c}
/k_{\rm B}$ provides a rough measure for the temperature below which
quantum effects due to nuclear tunneling become
important). For the special values $K=1/2$ and
$K=1$, the crossover behavior from high to low temperatures can
be solved explicitly \cite{egger93b}.

 In Fig.\ref{fig1} we show some results for
$P(t)$ in the high-temperature limit; in this parameter region,
 the effects of the initial
preparation are negligible, so $C(t)=P(t)$.
 For a study of the crossover
between nonadiabatic and adiabatic behaviors, all model
parameters except $\Delta$
are kept fixed. In the case of small $\Delta$,
the simulation results exhibit a monoexponential
decay on the golden rule timescale. However, for larger $\Delta$, the dynamics
becomes progressively more complex.
 After a fast initial transient, the decay slows down; fitting
this slower decay to an exponential law, one can again
extract a rate for this adiabatic situation.

The $\Delta$-dependent high-temperature rates measured in units of
$\omega_{\rm c}$ are plotted in Fig.\ref{fig2}. Clearly, the nonadiabatic
limit is nicely reproduced by the simulations for small $\Delta$,
 and the rates $\sim \Delta^2$. In the
adiabatic limit, the rate constant is approximately
{\em independent}\/ of the magnitude of the electronic coupling.
These results are in agreement with the conventional
Landau-Zener prediction (\ref{clas})
in the limit of high temperatures.
Remarkably, the data in Fig.\ref{fig2} show that
the golden rule formula is accurate for values of
$\Delta/\omega_{\rm c}$ as large as $\approx 1.5$. Note also that we find
a monotonic  dependence of the rate on $\Delta$, in
contrast to the findings of Skourtis et al.\cite{skourtis} which are based
on a rather simplistic Hamiltonian and would predict a maximum in the rate
as a function of $\Delta$.

Next we study the low-temperature region
 ($\hbar\beta\omega_{\rm c}= 2.5$), where the classical rate
formula (\ref{clas}) is not expected to hold.
Again, we find complete agreement
with the nonadiabatic golden rule formalism for
 small values of $\Delta/\omega_{\rm c}$, with
 $P(t)$ decaying  monoexponentially on the golden
rule timescale (with a fast initial transient).
 When increasing $\Delta/\omega_{\rm c}$ beyond
$\approx 0.5$, however, the dynamics exhibits an oscillatory behavior
as shown in Fig.\ref{fig3}. This prevents a meaningful estimate for the
decay constant (but when increasing $K$ to significantly larger values,
 we expect that the dynamics will become totally incoherent again, at least
for not too small $\omega_{\rm c}/\Delta$).
We note that in the limit $\omega_{\rm c}/\Delta \to 0$ (but $\lambda$
finite), the dynamics can be solved exactly.
This is the case of a strictly
 adiabatic bath \cite{chandler91}, where the dynamics
is always oscillatory.   From the simulations we also
observe that  the nonadiabatic regime
is confined to  increasingly smaller values for  $\Delta/\omega_{\rm c}$
when the temperature is lowered.

In conclusion, the simulations confirm the classical picture for
the crossover from nonadiabatic to adiabatic electron transfer
in the high-temperature limit and, in addition,
 provide a dynamical explanation.
In the low-temperature limit, however, the dynamics becomes
oscillatory unless the system-bath coupling is made very large.

\subsection{Antiferromagnetic Kondo region: Low-temperature dynamics}

We next turn to a different problem. The determination of the dynamical
quantities of interest in the low-temperature Kondo region characterized
by a Kondo parameter $1/2<K<1$ has been a long-standing problem.

We start with a brief discussion of the relationship of the spin-boson
model to
the Kondo problem\cite{tsvelick}.
The Kondo Hamiltonian in its simplest form
describes a spin-$\frac12$ impurity interacting with a band of
free electrons via isotropic exchange scattering.
A particularly useful method for studying the low-temperature spin
relaxation dynamics of the Kondo problem (the equilibrium properties
are well understood\cite{tsvelick}) employs a bosonization
procedure to map it onto a spin-boson problem with
Ohmic dissipation\cite{leggett87}.
 The case $1/2<K<1$ then corresponds to the interesting antiferromagnetic
Kondo problem. The important
dynamical quantity in the Kondo problem is the imaginary part
of the frequency-dependent
spin susceptibility, $\chi^{\prime\prime}(\omega)$.
It can be expressed in terms of the Fourier transform of the
correlation function (\ref{ctt})
\begin{equation}
C(\omega) = \hbar \coth(\beta\hbar\omega/2) \chi^{\prime\prime}(\omega)\;.
\end{equation}
Therefore a computation of $C(t)$ will yield
all relevant dynamical quantities (structure factor,
dynamical susceptibility, etc.) of the Kondo problem.

Unfortunately,
the noninteracting blip approximation (NIBA) by Leggett et al.\cite{leggett87}
cannot be justified in this region for temperatures below the Kondo
temperature defined as
 $T_{\rm K}= \hbar \Delta_r /k_{\rm B}$, where $\Delta_r=\Delta
(\Delta/\omega_{\rm c})^{K/(1-K)}$. In fact,
 if one equates $C(t)=P(t)$ (which NIBA predicts is true), the NIBA
would imply certain unphysical properties of the related Kondo problem,
such as a divergence in the susceptibility as $T\to 0$.

Furthermore, this
parameter region is also interesting in the context of
macroscopic quantum coherence, since there could
be remnants of coherent (oscillatory) behavior in the
zero-temperature dynamics for $1/2< K < 1$.
For this region, NIBA predicts a complete destruction of
quantum coherence.  We have studied $P(t)$ at
zero temperature in order to check this prediction.

Some numerical results for this parameter region have already been given
in Refs.\cite{mak92,egger92}, and a biexponential
behavior has been found. However, due to the limitations of
our earlier techniques, these
simulations were restricted to very small values
for the cutoff ($\omega_{\rm c} /\Delta=1.25$ has been used
in Ref.\cite{egger92}).
Furthermore, our previous algorithm did not allow for
a study of extremely low temperatures, and we have considered temperatures
only slightly below $T_{\rm K}$. These shortcomings can be overcome
by the algorithm discussed in Sec.~II, and we are now able to reach both the
scaling region $\omega_{\rm c}/\Delta\gg 1$ and the zero-temperature limit
for $P(t)$.
At zero temperature, the timescale of the dynamics should be
set solely by the frequency \cite{grabert}
\begin{equation}\label{deff}
\Delta_{\rm eff}  = [\cos (\pi K)\,  \Gamma(1-2K)]^{1/2(1-K)}\, \Delta_r
\end{equation}
where $\Gamma(x)$ denotes the Gamma function. This effective frequency
scale is equal to
 $\Delta$ at $K=0$, becomes equal to $\pi\Delta^2/2\omega_{\rm c}$
for $K=1/2$, and shrinks to zero as $K\to 1$.
Since the cutoff $\omega_{\rm c}$ should enter the dynamics
only via a renormalization of this effective frequency scale,
 we use the dimensionless time $y=\Delta_{\rm eff} t$; at zero
temperature, the only system parameter left is the Kondo parameter.
As shown by Grabert and Weiss \cite{grabert},
the NIBA solution for $T=0$ and $K<1$ can be written
in terms of the Mittag-Leffler function\cite{erdelyi},
\begin{equation} \label{mitlef}
P_{\rm NIBA}(y) = E_{2(1-K)} (-y^{2(1-K)})\;.
\end{equation}
In the parameter region $1/2\leq K <1$, this describes
a purely incoherent relaxation.

Before discussing the zero-temperature limit for $P(y)$ and $1/2<K<1$,
we first present data for the correlation function $C(t)$ at $K=1/2$.
This special case has been solved exactly by Sassetti and
Weiss \cite{sassetti}, and one finds
for sufficiently low temperatures that
$C(t)$ approaches zero {\em from below} at long times.
This is reproduced by
our simulations shown in Fig.\ref{fig4}; the cutoff chosen here
($\omega_{\rm c}=6\Delta$) is already large enough to ensure
the validity of $\omega_{\rm c}/\Delta\gg 1$. Note
that the corresponding exact solution for a factorizing initial state,
 $P(t)=\exp(-\Delta_{\rm eff} t)$,
 does not exhibit this behavior.

Finally, in Fig.\ref{fig5} we show QMC results for the
exact $T=0$ dynamics of $P(y)$ in the
Kondo region $1/2<K<1$. The cutoff chosen here is within the
scaling region $\omega_{\rm c}/\Delta\gg 1$ so that the dynamics
should depend only on the effective timescale $y=\Delta_{\rm eff} t$.
Indeed, as long as $\omega_{\rm c}/\Delta \geq 5$,
we found that a change in the cutoff enters
the dynamics solely via Eq.(\ref{deff}).
Since the frequency $\Delta_{\rm eff}$ becomes extremely small
with increasing $K$, it is not possible to
 numerically study the $T=0$ dynamics on timescales
of the order $\Delta_{\rm eff}^{-1}$ for $K$
larger than $\approx 0.75$. Here, we have restricted
ourselves to $K=0.6$ and $K=0.7$, see Fig.\ref{fig5}. For $K=0.5$, the
QMC data for $P(t)$ coincide with the exact solution.

It is not possible to fit the numerical data in Fig.\ref{fig5} to simple
(exponential, biexponential, algebraic, etc.) decay laws.  However,
it is obvious that {\em the NIBA gives the correct qualitative
picture}, especially at short times. The exact dynamics is
fully {\em incoherent}, yet of quite complicated appearance.
These results again underline our earlier finding
\cite{mak91,egger92} that {\em the NIBA provides an excellent
estimate for $P(t)$ in the bulk of parameter space}.
The data shown in Fig.\ref{fig5} suggest that the NIBA is
more accurate for $K=0.6$; this indicates that the NIBA
becomes exact for $P(t)$ as $K\to 1/2$.

One may question how much the respective correlation functions
$C(t)$ will deviate from the $P(t)$ depicted in Fig.\ref{fig5}, since
$C(t)$ is the relevant quantity for a comparison with the
Kondo problem.
While the NIBA was shown to give an excellent approximation
for the function $P(t)$ even in the low-temperature
Kondo region, the quality of the NIBA-prediction
$C(t)=P(t)$ seems to be poor in this region  (see
Fig.\ref{fig4}). We have carried out simulations
for $C(t)$ at temperatures slightly below
$T_K$ for $1/2 < K < 1$
as well, and the characteristic behavior shown in Fig.\ref{fig4},
namely $C(t)$ approaching zero from below
 as $t\to \infty$, was found
to persist. This would resolve the divergence of the
static spin susceptibility\cite{weiss93,sassetti}
predicted by NIBA based on $P(t)$ because $P(t)$ and $C(t)$
show qualitatively different behaviors.
Such a behavior of $C(t)$ is also in correspondence with
the exact Shiba relation\cite{shiba}.

Unfortunately, we were not able to reach the true zero-temperature
limit for the equilibrium correlation function $C(t)$.
 Clearly, using the same Kadanoff-Baym contour employed in
our method would require an infinitely long imaginary-time path
for $T=0$, which makes the algorithm impracticable.
By invoking ergodicity arguments \cite{weiss93}, however,
a viable variant of our algorithm may facilitate such
a calculation. To that purpose, one might consider a factorized
initial state at $t_0 <0$, so that for $t_0\to -\infty$
the system will be equilibrated at $t=0$. In effect, one is then
left with a real-time contour instead of the Kadanoff-Baym
contour, and the initial
correlations are represented by negative-time paths. This
method is currently under study.

\section{CONCLUSIONS}

We have proposed a real-time quantum Monte Carlo simulation
method for a numerically exact
computation of the dynamical quantities of the
spin-boson model. Our technique is based on a discretized
path integral formulation and makes use of the symmetry properties
of the dissipative influence functional, whereby one can integrate
out the quasiclassical paths and is left with a stochastic sampling
of the quantum fluctuations alone. This leads to a significant
improvement compared to earlier methods, and allows us to study
the dynamics at considerably longer times. The method is
generally applicable to dissipative tight-binding systems with arbitrary
spectral density.

Results have been presented for two problems of current interest.
(a) We have computed the electron transfer rate constant as a function
of the electronic coupling in both the high-temperature and
the low-temperature limit. In the high-temperature limit,
the classical Marcus result is reproduced; for low temperatures,
however, one obtains an oscillatory behavior for large electronic
couplings which does not allow for a simple rate description.
(b) The dynamics in the antiferromagnetic Kondo region
has been computed at $T=0 $ for several values of the
Kondo parameter. In accordance with the
 noninteracting blip approximation (NIBA),
we find a fully incoherent (yet non-exponential) decay
for the occupation probability $P(t)$. Initial
preparation effects, however, lead to important
deviations from the NIBA-prediction $C(t)=P(t)$ for the
equilibrium correlation function. It is exactly this
NIBA-equality which led to unphysical results for
the equivalent Kondo problem.

\acknowledgements

This work was partly supported by the  Camille and Henry
Dreyfus Foundation under the New Faculty Awards Program,
the National Science Foundation (CHE-9216221), and the
Young Investigator Awards Program (CHE-9257094). Computational
resources furnished by the IBM Corporation are acknowledged.
Finally, we wish to thank Uli Weiss for stimulating discussions.


\begin{table}
\caption[]{Performance of different dynamical simulation
techniques for spin-boson systems:
(a) a brute-force evaluation of the path integrals without
any filtering, (b) the SPMC
calculation \cite{mak91,egger92}, (c) the original discrete filtering
technique \cite{mak92,mak93a}, (d) the current optimized filtering method.
The data shown below are for a calculation of $P(t)$ for $K=1/2,
\omega_{\rm c}/\Delta = 6$,  both for $T=0$ and for a high
temperature, $\Delta \beta = 0.5$.
For times $t>t_{\rm max}$,
the dynamical sign problem becomes uncontrollable (with statistical
errors $> 20\%$).  }
\bigskip
\label{tab:I}
\begin{tabular}{l|c|c}
   & $\Delta t_{\rm max}$
for $\beta\Delta=\infty$ &
$\Delta t_{\rm max}$ for $\beta \Delta = 0.5$ \\ \hline
(a) & 5 & 8 \\
(b) & 9 & 14 \\
(c) & 12 & 18 \\
(d) & 22 & 24
\end{tabular}
\bigskip
\end{table}


\begin{figure}
\caption[]{\label{fig1}  Simulation results for $K=2,
 \, k_{\rm B} T/\hbar\omega_{\rm c} = 4$ and two values
of the electronic coupling. Note the change in timescale.}
\end{figure}

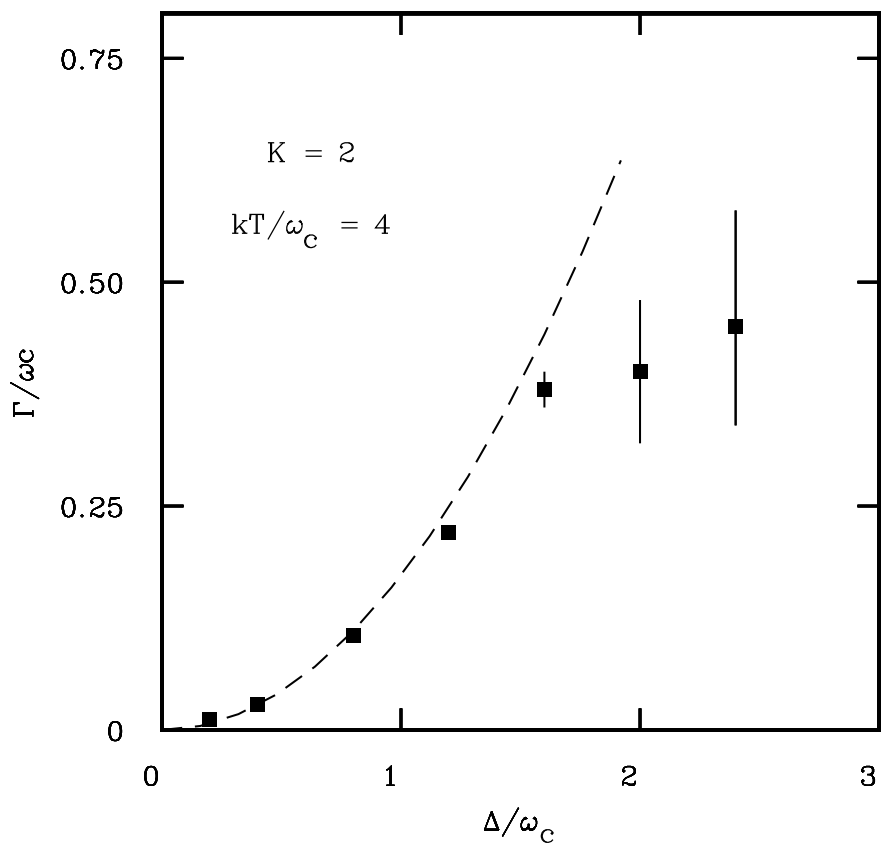
\begin{figure}
\caption[]{\label{fig2} Electron transfer rate constants as a function
of the electronic coupling; squares denote decay constants of $P(t)$
for $K=2,\, k_{\rm B}T/\hbar\omega_{\rm c} = 4$, i.e., total rates.
 The dashed line is the nonadiabatic golden rule prediction, and
 vertical bars are error estimates.}
\end{figure}

\begin{figure}
\caption[]{\label{fig3}  Simulation results for $K=2, \, k_{\rm B}T/
\hbar\omega_{\rm c} = 0.4$ and two values of the electronic coupling.
Note the change in timescale. }
\end{figure}

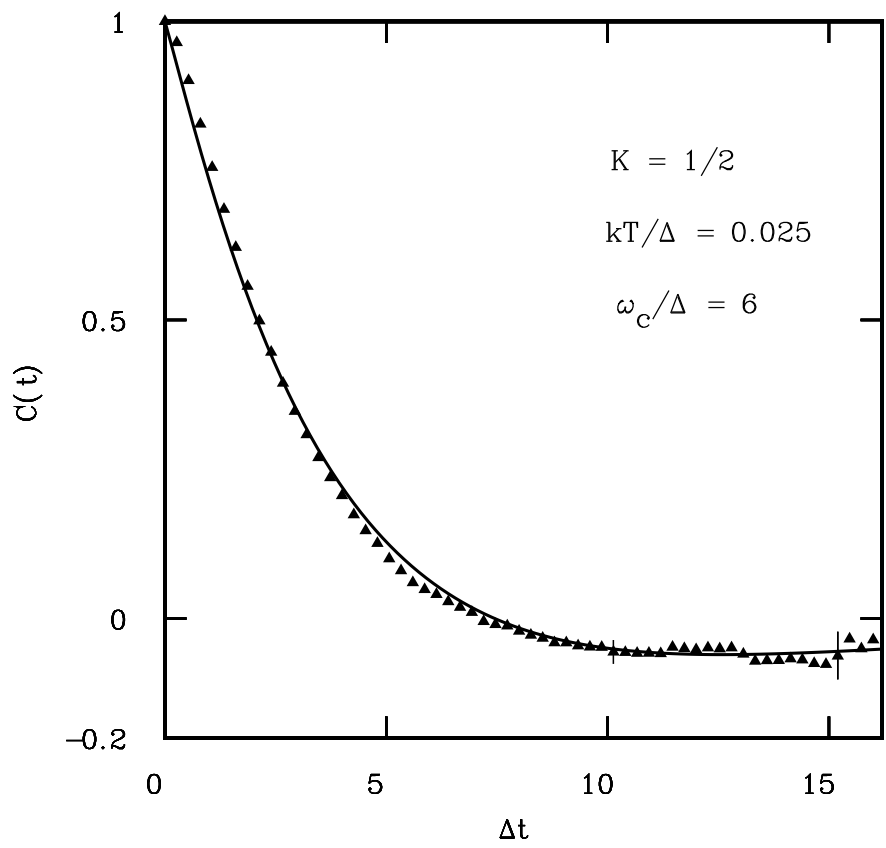
\begin{figure}
\caption[]{\label{fig4} Symmetrized correlation function $C(t)$ for
$K=1/2,\, \omega_{\rm c}/\Delta=6$ and $k_{\rm B} T/\hbar\Delta
= 0.025$. The triangles denote the QMC data for
discretization numbers $q=60$ and $r=130$. The solid curve is the exact
solution from Ref.\cite{sassetti}, and vertical bars are error estimates.}
\end{figure}

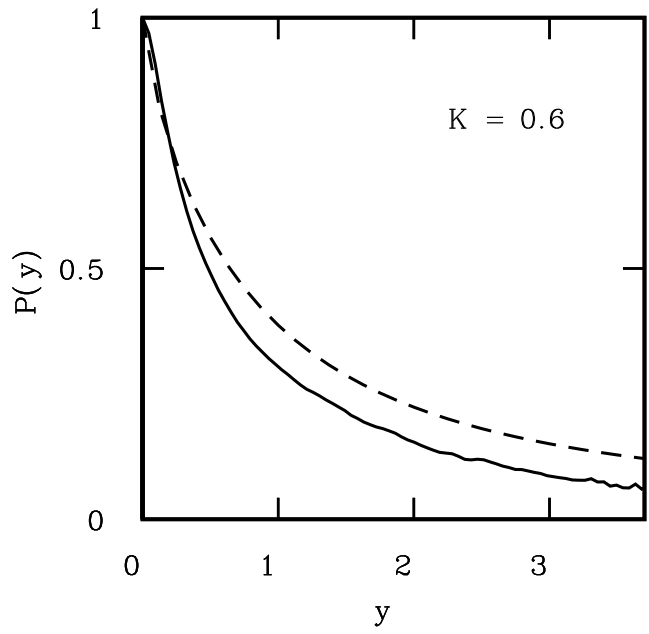
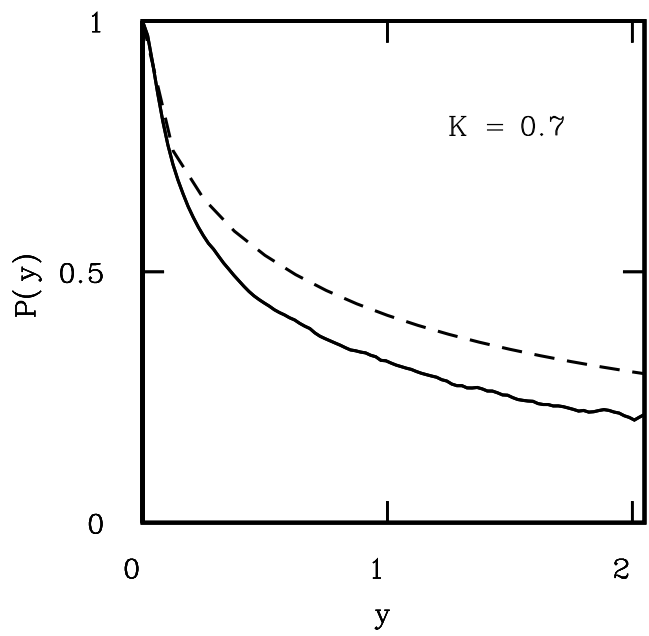
\begin{figure}
\caption[]{\label{fig5}
 Zero-temperature dynamics of $P(y=\Delta_{\rm eff} t)$
 in the antiferromagnetic Kondo region.
 The solid curves are numerically exact results,
 dashed curves are NIBA predictions. Note the change in
 effective timescale between both plots.
 The discretization numbers were $q=80$ and $q=160$ for
 $K=0.6$ and $K=0.7$, respectively. }
\end{figure}


\begin{references}
\bibitem{foot1} The last term in the influence functional
(\ref{inflma}) modifies only the matrix ${\bf V}^{(1)}$. Note that the
path combinatorics is already contained in the definition of the
${\bf V}^{(j)}$ matrices, since some of these matrix elements are zero.
Finally, for a general tight-binding system
 with $d$ states, we have to deal with
a matrix multiplication of $q\;\, (2d+1 \times 2d+1)$ matrices.
\end{references}
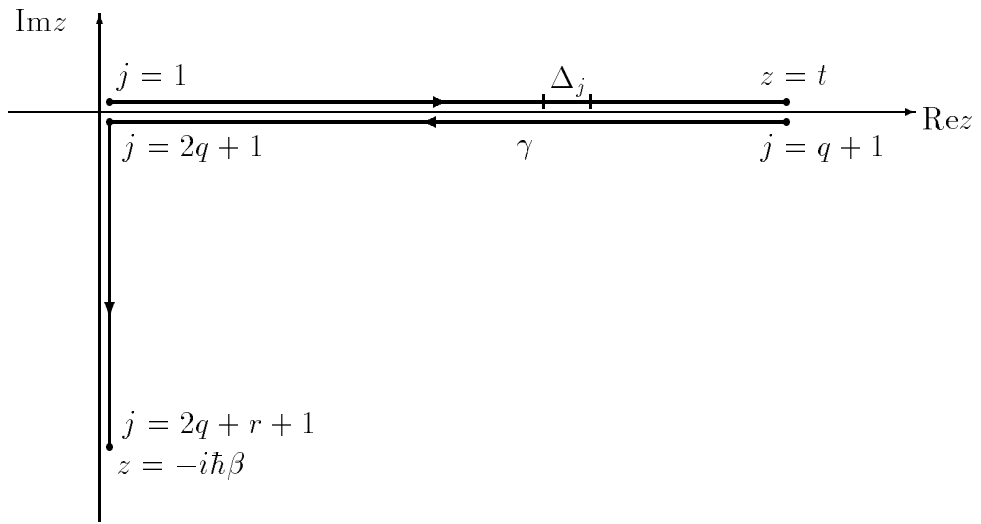
\end{document}